# Oxidation of Monolayer WS$_2$ in Ambient is a

# Photoinduced Process.

Jimmy C. Kotsakidis[1*], Quianhui Zhang[2], Amadeo L. Vazquez de Parga[3,4], Marc Currie[5], Kristian Helmerson[1], D. Kurt Gaskill[5], Michael S. Fuhrer[1*]

[1] School of Physics and Astronomy, Monash University, Victoria 3800, Australia.

[2] Department of Civil Engineering, Monash University, Victoria 3800, Australia.

[3] Dep. Física de la Materia Condensada and Condensed Matter Physics Center (IFIMAC), Universidad

Autónoma de Madrid, Cantoblanco 28049, Madrid, Spain.

[4] IMDEA Nanociencia, Cantoblanco 28049, Madrid, Spain.

[5] U.S. Naval Research Laboratory, Washington D.C. 20375, USA.

*Address correspondence to Michael.Fuhrer@monash.edu or Jimmy.Kotsakidis@monash.edu

**Abstract**

We have studied the ambient air oxidation of chemical vapor deposition (CVD) grown monolayers of the semiconducting transition metal dichalcogenide (S-TMD) $WS_2$ using optical microscopy, laser scanning confocal microscopy (LSCM), photoluminescence (PL) spectroscopy, and atomic force microscopy (AFM). Monolayer $WS_2$ exposed to ambient conditions in the presence of light (typical laboratory ambient light for weeks, or typical PL spectroscopy map), exhibits damage due to oxidation which can be detected with the LSCM and AFM; though may not be evident in conventional optical microscopy due to poorer contrast and resolution. Additionally, this oxidation was not random, and correlated with 'high-symmetry' and red-shifted areas in the PL spectroscopy map – areas thought to contain a higher concentration of sulfur vacancies. In contrast, samples kept in ambient and darkness showed no signs of oxidation for up to 10 months. Low-irradiance/fluence experiments showed that samples subjected to excitation energies at or above the trion excitation energy (532 nm/2.33 eV and 660 nm/1.88 eV) oxidized in as little as 7 days, even for irradiances and fluences eight and four orders of magnitude lower (respectively) than previously reported. No significant oxidation was observed for 760 nm/1.63 eV light exposure, which lies below the trion excitation energy in $WS_2$. The strong wavelength dependence and apparent lack of irradiance dependence suggests that ambient oxidation of $WS_2$ is initiated by photon-mediated electronic band transitions, that is, photo-oxidation. These findings have important implications for prior, present and future studies concerning S-TMDs measured, stored or manipulated in ambient conditions.

Since the discovery of two-dimensional (2D) van der Waals materials such as graphene[1] and the semiconducting transition metal dichalcogenides (S-TMDs),[2] 2D physics has become more accessible to laboratories around the world, leading to an exponential increase of published papers year upon year.[3-4] S-TMDs (and similarly graphene) possess radically different and useful properties in their 2D, monolayer form. Properties shared among some of the monolayer S-TMDs of family (Mo, W)(S, Se)$_2$ include a large direct bandgap,[5-7] exceptional optical characteristics[8] and strong spin-orbit coupling along with broken inversion symmetry.[9] This has lead various groups to demonstrate the usefulness of S-TMDs in potential applications ranging from ultra-low power electronics,[10] valley-tronics,[11-12] photonics[13] and qubits[14] to gas sensors.[15]

Understanding S-TMDs stability in ambient conditions and under light illumination – crucial for measurements and manipulations undertaken in those conditions – is essential for their development into potential applications. It is now well-known that monolayer and multilayer S-TMDs oxidize upon exposure to extreme conditions such as ultra-violet (UV) light irradiation in moisture-rich conditions[16-17] or elevated temperatures in ambient atmosphere.[18-21] More recently however, monolayer S-TMD oxidation has been reported in ambient conditions. For example, Gao *et al.*[22] were among the first to report that WS$_2$ and MoS$_2$ had poor long-term stability in ambient conditions and observe that H$_2$O greatly sped up the oxidation process. Gao suggested the oxidation began with oxygen substitution at sulfur vacancy sites and progressed *via* O or OH radicals, with H$_2$O acting as a 'catalyst' lowering the energy barrier for reaction. In a subsequent report, Kang *et al.*[23] showed that oxidation was greatly suppressed, though not eliminated, when the S-TMD (in this case WS$_2$) was epitaxially grown on graphene compared to an oxide substrate. It was proposed that electric fields were necessary for WS$_2$ oxidation, although the detailed mechanism was unclear. Recently, Atkin et al.[24] determined that the oxidation reaction could be initiated by laser light (440 nm), and proposed a fluence threshold (>1.5×10$^{10}$ J m$^{-2}$) necessary for oxidation. Atkin also found that $H_2O$ was necessary for the oxidation reaction to proceed at any measurable rate (in agreement with Gao),[22] and in addition, found that sulfate was a likely reaction product.[24] Although these studies have identified important factors in the oxidation process *i.e.* ambient conditions, humidity, substrate, reaction products and radiant exposure levels – a complete and fundamental understanding of the conditions under which oxidation takes place, and more critically the conditions that completely avoid oxidation, is lacking.

In this work, we investigate ambient-exposed monolayers of the S-TMD $WS_2$ using standard characterization tools – optical microscopy, photoluminescence (PL) spectroscopy and atomic force microscopy (AFM). Moreover, we also employ laser scanning confocal microscopy (LSCM), which serves as an integral tool in this work. By correlating AFM images with LSCM images, we are able to rapidly identify oxidized regions of monolayer $WS_2$, which may not be evident from optical microscopy or PL spectroscopy. Oxidation is evident in samples kept in ambient conditions and exposed to light characteristic of laboratory conditions (typical room light for weeks), or moderate laser powers in PL mapping spectroscopy, but is not seen in as-grown samples or those stored in ambient conditions and darkness for up to ten months. We further explore the role of light by irradiating monolayer $WS_2$ samples with low-irradiance, visible light at 532, 660, and 760 nm. For $WS_2$ samples exposed to 532 and 660 nm light (above the threshold necessary for electronic excitation), oxidation was observed, whereas, samples exposed to 760 nm light did not show appreciable oxidation (similar to those left in darkness). These results indicate that oxidation of $WS_2$ in ambient conditions requires photo-excitation, *i.e.*, oxidation is a wavelength dependent, photo-induced process – at odds with recent interpretations in literature.[22-24] Furthermore, photo-induced oxidation occurs at irradiances and fluences of eight and four orders of magnitude lower, respectively, than the irradiance and fluence thresholds proposed in Ref. 24; suggesting that the threshold for oxidation is extraordinarily low or absent. To our knowledge, a photo-oxidation mechanism in ambient conditions has not been previously described in detail for any member of the 2D S-TMD family of materials. The details of the oxidation reaction – likely Förster resonance energy transfer (FRET) and/or photo-catalysis involving redox reactions with $H_2O$ and $O_2$ – requires further study. However, we expect the ambient air photo-oxidation of other direct bandgap S-TMDs (such as $MoS_2$) to occur similarly. These findings are important for future S-TMD based optoelectronic and electronic applications, since these findings establish important protocols for all researchers seeking to avoid damaging S-TMDs *via* oxidation.

## Results

WS$_2$ is grown on c-sapphire (single side polished) in an atmospheric-pressure chemical vapor deposition (CVD) process described previously,[25] using mixtures of argon/hydrogen with $WO_3$ and sulfur powder precursors (see Methods and Figure S1 in the Supporting Information for details).

Figure 1a shows an LSCM micrograph of an as-grown monolayer WS$_2$ crystal, stored in darkness after growth for approximately one month, and exposed to laboratory and LSCM light for short durations before being stored again in darkness. Figure 1b shows LSCM and optical micrographs taken on a similarly grown sample after approximately 19 days of exposure to ambient atmosphere and typical laboratory room lighting. The LSCM image (left side of Figure 1b) shows dark spots within the $WS_2$ crystal in Figure 1b, although these are not apparent in Figure 1a. These dark spots are difficult or impossible to observe in the optical micrograph of the same crystal, shown on the right side of Figure 1b (and similarly in Figure 1c). As discussed in more detail below, the dark spots correspond to the complete or partial oxidation of small (< 0.5 μm) regions of $WS_2$ (*i.e.* $WO_x$, where $x \leq 3$). This observation is consistent with recent reports on the ambient air oxidation of monolayer S-TMDs, which have recorded samples in laboratory conditions oxidizing in months[22] to as little as weeks.[23] After the observation of dark spots/oxidation in Figure 1b, care was taken to not expose freshly grown samples to ambient UV light in the room, as this was thought to contribute to the oxidation of the $WS_2$.[16-17] Thus, freshly grown crystals were stored in darkness. To understand the optical properties of our CVD grown $WS_2$ and investigate whether this could provide clues as to the causes of the observed oxidation in Figure 1b, samples were mapped using confocal PL spectroscopy (μ-PL). Figure 1c shows LSCM and optical micrographs of a monolayer $WS_2$ crystal after performing μ-PL of the sample, but otherwise the sample was protected from light, while Figure 1d shows a sister sample imaged with the LSCM after storage in darkness for 10 months. The conditions of the μ-PL imaging (excitation wavelength 532 nm, power 140 μW, irradiance $\approx 6.9 \times 10^8$ W m$^{-2}$, with total fluence of $\approx 1.8 \times 10^8$ J m$^{-2}$, see Methods for details) are similar to those routinely used for optical characterization of S-TMD crystals.[26] Table 1 shows all the irradiances and fluences used in this report. For the PL-exposed sample in Figure 1c, similar dark spots as in Figure 1b are evident with the LSCM micrographs, though are not obvious in the conventional optical micrograph shown on the right side of Figure 1c. The standard sample shown in Figure 1d, which remained in darkness, appears pristine (in Figure 1d, a bright spot in the middle of the $WS_2$ is evident; we identify this feature as a crystal seed center or multi-layer $WS_2$, and not associated with oxidation). The results imply that light is responsible for the ambient air oxidation of monolayer $WS_2$. Considering there are many reports using PL excitation conditions of S-TMDs on the same order of magnitude as that used in this report, the observation of oxidation after μ-PL was unexpected. Moreover, since the LSCM uses light to image the samples, it also carries a probability of oxidizing the $WS_2$. Thus, care was taken to ensure that the final step in the analysis was light exposure from the LSCM, and that sister samples (such as the one in Figure 1d) were used as standards to compare with the effects of any analysis/light exposure experiments that were undertaken. Yet even after two subsequent LSCM exposures, no oxidation is observed, the details of which can be found in the Supporting Information (Section 2, Figure S2). As will become clear, the advantages of LSCM characterization stem not only from its superior contrast and resolution, but also from its inherent operation at low incident fluence – much lower than those found in PL experiments, or under low-light illumination for days. The LSCM can operate in two modes, the first uses white light from an LED (optical micrographs in Figures 1b and 1c) and the second uses 405 nm light in a confocal laser scanning setup. In the case of the white light, the $WS_2$ is subjected to an irradiance of $\approx 1.6 \times 10^4$ W m$^{-2}$ and fluence of $\approx 1.6 \times 10^5$ J m$^{-2}$ (for 10 second exposures at 100× magnification). In the case of the 405 nm light, the irradiance and fluence were $\approx 1.3 \times 10^8$ W m$^{-2}$, and $\approx 8.8 \times 10^4$ J m$^{-2}$, respectively (see methods for details).

Figures 2a-d show the results of μ-PL on a different crystal on the same sample as in Figure 1c. Figure 2a shows the μ-PL intensity distribution of the $WS_2$ monolayer. It can be seen that the PL intensity on the edges is much brighter than that of the center, as is observed in many other reports.[6, 24, 26-31] Also, the μ-PL shows a three-fold rotationally symmetric pattern, not only in the intensity map, but also mirrored in the map of the peak photon energy (relating to the location of the exciton) shown in Figure 2b, *i.e.*, the two correlate with each other.[28] Figure 2c shows a representative spectrum from the central region of the crystal, deconvolved as two Voigt[32] spectra identified as the exciton peak at approximately 2.03±0.01 eV and smaller trion peak at 1.97±0.01 eV. A representative spectrum from the edge of the crystal is shown in Figure 2d, and is similarly deconvolved, showing a shifted exciton peak at approximately 1.96±0.01 eV and trion peak at 1.91±0.01 eV. Figures 2c and 2d show the heterogeneity of the optical characteristics of CVD grown monolayer $WS_2$ on sapphire; similar optical heterogeneity is found in all other samples. After μ-PL analysis, the sample was stored in darkness and then imaged with the LSCM as shown in Figure 2e. As in Figure 1c, the sample shows signs of oxidation, manifesting as small (<0.5 μm) dark spots, similar to those observed in previous reports using conventional light microscopy or atomic force microscopy.[17, 20-23, 26]

To better understand the oxidized regions seen with the LSCM, the same sample was then imaged with AFM in intermittent-contact mode, shown in Figures 2f-h. Figure 2f shows an overview of the sample, and the extent of the oxidation. AFM resolves more clearly the dark spots seen in LSCM as raised triangular 'islands'. These triangular islands in S-TMDs have been observed before and identified as tungsten oxide in various oxidation states, $WO_x$ ($x \leq 3$).[17, 23-24, 26] We see that upon comparison of the μ-PL shown in Figures 2a and 2b and the after effects of oxidation seen in the LSCM and AFM micrographs of Figures 2e and 2f respectively, that oxidation occurs more prevalently in some regions in a well-defined pattern – following the three-fold symmetric pattern of lines extending from the center towards the vertices of the triangular crystal. Also, significant oxidation can be seen outside these areas, for example in the region with red-shifted PL corresponding to the spectrum in Figure 2d. The inset of Figure 2f shows the height of the central part of the crystal as ≈ 0.5 nm, with the very edge of the crystal higher than that of the central part by $\approx 0.5$ nm, with edge width of $\approx 0.32$ nm (see Supporting Information Figure S5). As can be seen in the inset of Figure 2f (see also Figure S5), the edge is 'granular' and contains local increases in height some of which are above or below 0.5 nm, the maximum of which was $\approx 1.3$ nm higher than the $WS_2$ basal plane (see Figure S5 in Supporting Information section 4). In some samples (from different growth runs), this edge was also discernable under the LSCM (see Supporting Information Figure S4). This higher edge region correlates well to the increased edge PL intensity of the μ-PL map (Figure 2a), and thus we tentatively ascribe the bright edge in the PL spectrum to this physical feature observed with the AFM. Increased PL brightness around the edges of monolayer S-TMDs has also been reported on by others,[6, 24, 26-31, 33] and the effect is attributed to either differences in chemical composition on the edges of the crystal[6, 26-30, 33] or to water intercalation at the edges of the S-TMDs.[24, 31, 34]

Figure 2g and 2h show higher-resolution AFM topography and phase images respectively, of the triangular oxidation island feature outlined by the dotted black box in Figure 2f. The AFM topography image in Figure 2g shows that these triangular oxidation islands are not holes (as suggested by the LSCM in Figure 2e) but raised in topography by $\approx 1.14$ nm above the surrounding $WS_2$; this is consistent with thicker $WO_x$ remaining as the reaction product (see Supporting Information Section 3 for in-depth analysis of oxidation heights). The phase contrast between the oxidation island and $WS_2$ (Figure 2h) also indicates that the oxidized area is a different material. Thin films of $WO_x$ have a much higher transmittance than $WS_2$ at visible wavelengths (see Supporting Information Figures S4 and S5 for transmittance, reflectance and absorptance of monolayer $WS_2$) explaining the dark (see-through to substrate) appearance of these regions under the LSCM.[35] It was observed that for some oxidation islands, there was a small (<60 nm), round, raised region in the middle of the oxidation island. This spot, seen more clearly with the enhanced contrast of the phase image in Figure 2h, suggests this feature is a liquid drop, and could result from hygroscopic sulfur oxides.[24]

The experiments thus far imply that visible light with wavelengths as long as 532 nm can cause oxidation of $WS_2$ in ambient, at irradiances typically used for PL spectroscopy. To determine the physical mechanism of $WS_2$ oxidation, controlled, low-irradiance light exposure experiments were conducted with monolayer $WS_2$ samples at three specific wavelengths. The low irradiances used allow us to completely eliminate heating as a source of oxidation and focus on the wavelength dependence of the oxidation. The experiments used two lasers at 532 nm (2.33 eV) and 660 nm (1.88 eV) (see Methods for details and Supporting Information for laser spectra) and a halogen light (XGY-II) filtered at 760 nm (1.63 eV), with irradiances of approximately 3.8, 14.2, and 2.2 W m$^{-2}$, respectively. Previous experiments observed heating only at irradiances approximately nine orders of magnitude higher (in $MoS_2$).[36] Figure 3a shows the absorptance spectrum of the $WS_2$ monolayers at ≈293K, determined *via* subtracting the measured transmission and reflectance spectra from the total light impinging on the sample (additional details in Methods). The absorptance spectrum is consistent with previous observations for monolayer $WS_2$,[37-39] and shows features characteristic of the 'A' exciton at 619±2 nm (2.003±0.006 eV) and 'B' exciton at 516±2 nm (2.403±0.009 eV), which arise from excitonic transitions in the spin-orbit split bands at K/K' points in the Brillouin zone. The observed spin-orbit splitting, $\Delta_{SO}$, of 0.400±0.015 eV is close to previously measured values for monolayer $WS_2$.[38]

The corresponding wavelengths used in the controlled low-irradiance light exposure experiments are highlighted and superimposed on Figure 3a. The 532 nm light was chosen to mimic the conditions found in the PL experiment, the 660 nm light was chosen as it is at the edge of the absorption of $WS_2$ (excites mainly the trion) and 760 nm light which is not significantly absorbed by the $WS_2$ (*i.e.* photon induced electronic transitions are improbable). Light was shone constantly on the samples for 7 days in ambient atmosphere (ranging from $\approx 40\%$ to $\approx 60\%$ humidity) and they were subsequently imaged using the LSCM, giving a total fluence (un-corrected for absorptance) of $2.3 \times 10^6$, $8.6 \times 10^6$ and $1.3 \times 10^6$ J m$^{-2}$ for the 532, 660 and 760 nm light respectively. Note that these fluences are at least an order of magnitude larger than that of the LSCM.

Figures 3b-e show the results of the low-irradiance experiments after 7 days of exposure to the different wavelengths (b-d) or kept in darkness for two weeks (e) as imaged with the LSCM. In Figures 3b and 3c, small triangular islands on the edges and the interior of the $WS_2$ indicate that the 532 nm (b), and 660 nm (c) light caused significant oxidation of these samples. Figure 3d shows the results of 760 nm light exposure; no significant amount of oxidation can be detected, though on some crystals, minute traces of oxidation is observed (see Section 9 in Supporting Information). All samples (including those in Figures 3b to 3d) were exposed to small amounts of ambient light while mounting/removing the samples from the optical setup (and CVD furnace) and loading/measuring the samples for analysis with the LSCM. Additionally, the 760 nm exposure used a filter (see Methods) instead of a pure wavelength source such as a laser, which may have further contributed to stray light exposure. As can be seen in Figure 3d, these brief ambient light exposures do not cause extensive oxidation (as observed with the LSCM) compared to Figures 3b

and 3c which are severely oxidized. Figure 3e shows an LSCM image of a control sample kept in darkness for two weeks after growth; no oxidation in any samples kept in darkness was observed.

We also exposed the $WS_2$ monolayers to similar low-irradiance ($\approx 3$ W m$^{-2}$) green ($\approx 515$ nm) light in a nitrogen filled glove box environment with ultra-low levels of $O_2$ and $H_2O$ for 7 days. In contrast to the ultra-low irradiance experiments in ambient conditions, the results show that no damage or adverse effects were observed (using the LSCM) on glove box exposed $WS_2$ (see section 10 in Supporting Information for details); indicating that light-induced oxidation does not proceed in the absence of $O_2$ and $H_2O$.

## Discussion

We have identified for the first time that the previously observed[22-24] ambient oxidation of a monolayer S-TMD requires exposure to light that is able to excite electronic transitions within the $WS_2$, *i.e.*, the physical mechanism of oxidation in ambient is photo-oxidation. While heating of TMDs in ambient has been shown to cause oxidation in S-TMDs,[18-22] we can rule out heating as the cause of oxidation in our experiments, as we observe similar oxidation for long-time exposures at low-irradiance in Figures 3b and 3c. These low-irradiances are approximately eight orders of magnitude lower than the high-irradiance μ-PL exposure (Figures 1c and 2) and are much below any levels that would cause measurable heating.[36] Furthermore, the LSCM uses higher irradiance than the green or red low-irradiance experiments, yet results in no significant oxidation. This implies that irradiance is not an important factor for oxidation. In contrast, fluence does appear to be essential for oxidation. Light exposure at 440 nm (2.82 eV) at high fluence ($>1.5 \times 10^{10}$ J m$^{-2}$) has been previously observed to cause oxidation of monolayer S-TMDs.[24] Here we find that lower photon energy light (down to the threshold for electronic excitation, 1.88 eV or 660 nm in WS$_2$) is sufficient for oxidation, using a fluence approximately 4 orders of magnitude lower ($\approx 2.3 \times 10^6$ J m$^{-2}$ in Figure 3b) than previously reported. It should be noted, that by taking into account the absorptance of WS$_2$ (from Figure 3a), the actual amount of absorbed light (*i.e.* light that can excite electronic transitions) is lower by an order of magnitude for the 532 and 660 nm light and approximately zero for the 760 nm light (the uncorrected values of irradiance and fluence are quoted here for direct comparison with literature, see Supporting Information section 6). In comparison, the LSCM images, which show no obvious signs of oxidation, were obtained using fluences of $\approx 8.8 \times 10^4$ J m$^{-2}$. A summary of the irradiances, fluences and whether oxidation was visible in each light exposure experiment is presented in Table 1. These observations suggest that a fluence threshold for oxidation is either non-existent or extraordinarily low, and that oxidation will simply proceed at a rate that is proportional to the fluence and the efficacy in which the material generates excited carriers from optical excitation. Thus, depending on the resolution of the probe used, oxidation may not be observed, yet still be present even with brief exposures at low fluence.

The strong wavelength dependence of oxidation indicates a photo-excitation mechanism involving the WS$_2$ itself. We can rule out direct excitation of oxygen to create singlet oxygen (excited, reactive form of oxygen – see Supporting Information), which would occur at $760 \pm 10$ nm wavelength excitation,[40-41] as we observe no oxidation in the low-irradiance 760 nm/1.63 eV experiment. The fact that the threshold photon energy (wavelength) corresponds well to the energy (wavelength) threshold for electronic excitation (excitation of the trion), strongly suggests that oxidation instead occurs through a photo-excitation process, *i.e.* photo-oxidation. The detailed mechanism — likely Förster resonance energy transfer (FRET) and/or photo-catalysis reactions involving excited $H_2O$ and/or $O_2$ species — requires further study (see Supporting Information Section 7 for more details). However, we also expect the ambient air photo-oxidation of other direct bandgap S-TMDs (such as $MoS_2$) to occur similarly due to very similar chemistry. For instance, Ding *et al.*[42] have recently discovered singlet oxygen present in photo-excited solutions of $MoS_2$ quantum dots, and have suggested FRET as the mechanism of oxidation. The necessity of photo-excited carriers for oxidation also explains the protection from oxidation of monolayer S-TMDs placed on graphene:[23] semi-metallic graphene efficiently quenches the photoexcited carriers through non-radiative recombination, thus greatly reducing the available excited carriers for chemical reaction *via* FRET and/or photo-catalysis.[43]

We note at this point that the photo-oxidation of semiconductors (especially direct bandgap) is not a new observation, and has been reported on since at least 1977 in materials such as GaAs[44-46] (see reference 45 and references therein) and with other direct bandgap semiconductors such as InP[46] and $TiO_2$.[47] Other 2D materials such as black phosphorous and the recently discovered 2D ferromagnet, $CrI_3$, have also been found to photo-oxidize (more rapidly compared to $WS_2$) when exposed to ambient conditions and light.[48-49] In the ambient air photo-oxidation of black phosphorous and $CrI_3$, $H_2O$ has been identified as a crucial reactant for oxidation. In the case of black phosphorous, oxidation required the presence of both adsorbed water and oxygen – oxidation did not occur unless both molecules were present, even though the oxidation rate was found to only depend on the concentration of $O_2$.[48] This was also found with GaAs, in which $H_2O$ was found to have a 'catalytic' effect on oxidation.[50] S-TMDs are no exception, and have also been found to only oxidize substantially in ambient with $H_2O$ present,[22, 24, 51-53] suggesting a similar mechanism may be responsible. The photo-oxidation of semiconductors is not limited to direct bandgap materials, and has also been reported for silicon (note that $CrI_3$ also has an indirect bandgap),[54] although the effect was found to be less pronounced than for direct bandgap materials.[46] Thus, it is expected that even multilayers of S-TMDs will undergo ambient air photo-oxidation, and indeed this has been reported by Budania *et al.*[52] which observed oxidation in exfoliated multilayer $MoS_2$ samples left in ambient conditions (though the role of light was not investigated). The photo-induced oxidation of semiconductors in ambient atmosphere seems to be a universal effect, yet not much attention has been paid to the mechanism in the 2D community, especially in regard to S-TMDs. Thus, we expect that past experiments may have been affected by photo-oxidation (see Supplementary Information Section 8) since our results have shown that even for ultra-low light exposures (see Table 1), that it only takes $\approx$ days for there to be significant oxidation present. For example, oxidation may have affected past S-TMD electrical devices by increasing contact resistivity ($WO_x$ has a larger bandgap), explaining at least in-part the large variation in measurements and lack of reproducibility from group to group and device to device.[55]

Furthermore, we observe that oxidation does not occur randomly across individual monolayer $WS_2$ crystals, but rather, shows a preference for regions running roughly from the centers to the vertices of triangular crystals, *i.e.* arranged with threefold rotational symmetry around the crystal centers (Figure 2e-f), as well as the edges and red-shifted PL regions of the crystals (Figures 1b, 3b, 3c, S10, S11b and Figures 2a, b and d respectively). The same threefold rotationally symmetric pattern is seen in the μ-PL (Figure 2a-b) and has been observed in our group previously,[31] and by other groups.[26-30] Recent work counting individual defects using conductive AFM has demonstrated that this three-fold symmetric region has a higher defect density than surrounding areas of the single crystal.[56] Additionally, this three-fold symmetric region has previously been shown to be preferentially oxidized under high-power laser irradiation and suggests that it contains a higher density of defects.[26] The observation of increased defect density has not been limited to the three-fold symmetric areas. Carozo *et al.*[57] measured defect densities in CVD grown $WS_2$ using scanning transmission electron microscopy and found that the edges of single crystals typically have higher sulfur vacancies than the interior of the crystal. More recently, Hu *et al.*[33] analyzed CVD-grown $WS_2$ using PL and has found that the bright edges observed in PL are most likely due to oxygen chemically bonded to the edges. In regards to the red-shifted PL regions, it has been demonstrated that short duration plasma treatments can increase defect density in $WS_2$, creating a neutral exciton peak that is red-shifted in energy $\approx 0.1$ eV below the 'defect free' neutral exciton peak at $\approx 2.03$ eV (see Figure 2c).[32] In that study, high resolution transmission electron microscopy was performed and in conjunction with simulations, suggested that the red-shifted neutral exciton peak was due to single sulfur vacancies. These prior results suggest that the oxidation we observed in our experiments may have begun at three-fold symmetric regions, edge sites and red-shifted regions due to the increased defect densities contained there.

There is now overwhelming evidence that the most common structural defects present in S-TMDs are in the form of sulfur vacancies, and this is true (with the exception of atomic layer deposition grown samples)[58] whether the samples are exfoliated,[58-60] grown *via* chemical vapor transport (CVT)[61] or, as in our case, CVD.[30, 56-58] When sulfur vacancies are present on the surface of $WS_2$, oxygen dissociative chemisorption in the these sites has been experimentally[62] and theoretically[63] shown to be favorable. Thus, by forming the initial seed center, it is likely that sulfur vacancies chemisorbed with oxygen nucleate the oxidative growth of $WO_x$ species.

In addition to the small (< 250 nm) triangular oxide islands formed (Figure 2e-f), we observed with the AFM small (< 65 nm) droplets in the centers of some of the triangular oxidation islands. The contents of these droplets are most likely aqueous $H_2SO_4$ ($H_2SO_4$ is hygroscopic) which forms one of the products in the oxidation of S-TMDs.[24, 51, 53, 64-65] Studies on the oxidation of $MoS_2$ have found traces of molybdic acid ($MoO_3 \cdot H_2O$ and $MoO_{2.4} \cdot xH_2O$, x = 0.7 − 1),[51, 65] though due to the low to negligible solubility of tungstic and molybdic oxides in water (or $H_2SO_4$),[65] it is unlikely that these droplets contain any significant quantities of tungstic acid. Not all triangular oxidation islands were observed to have a droplet in the middle of them, and this could be due to tungsten oxide forming in the absence of excess water.

However, the physical reasons behind the changes in PL intensity[26-31] and peak position[28] observed in the PL of $WS_2$ (Figures 2a-d) remain unclear. In particular, some groups have considered strain from lattice mismatch between crystal and substrate to explain changes in PL, but have subsequently ruled out this strain due to the persistence of the threefold symmetric pattern upon transfer of the crystal to another substrate, which is thought to relieve strain in as-grown crystals.[26, 29] Since oxidation was also observed in a region in which the exciton (and trion) emission peak was red-shifted (see Figures 2a-d), it may be that our measured red-shifted peak has contributions from both the trion and a defected neutral exciton (due to sulfur vacancies); further studies are required for confirmation. What is clear, is that our results indicate that the process of obtaining

PL spectra in ambient conditions will necessarily form small oxide regions and hence, the analysis of the spectra could be affected by several associated mechanisms, for example, strain imparted to surrounding $WS_2$ developed during oxide formation.[66] But there is reason to believe this issue can be mitigated as we have shown that illumination by light in a nitrogen atmosphere results in no measurable oxidation (see Supporting Information Section 10). Thus, performing PL in an inert environment is likely to be useful. Further work is needed, for example by growing crystals and performing PL in an inert environment or vacuum, and then exposing these crystals to ambient and light in order to determine the effect of defected regions on PL. Additional work is also needed to confirm the nature of the oxidation mechanism (FRET or photocatalysis) and whether defects are necessary to initiate the oxidation, or whether it can proceed in a defect-free basal-plane of $WS_2$ in ambient with light exposure. Further elucidation of the oxidation mechanism would enable accurate prediction of the oxidation rate, which would be valuable in in other aspects of S-TMD research and not limited to future oxidation studies.

## Conclusion

Our work shows that the physical mechanism that causes the oxidation of $WS_2$ in ambient conditions is a photoinduced, *i.e.*, a photo-oxidation mechanism. Oxidation is observed to not occur randomly, but in three-fold symmetric areas (as defined by μ-PL), edges and areas in which there is an observed red-shift in exciton emission. It is thought that all these areas contain more defects in the form of sulfur vacancies where oxygen can attach, and so provide a starting point from which oxidation can proceed. The fluence threshold that begins this photo-oxidation is at least four orders of magnitude lower than previously thought,[24] $<2.3 \times 10^6$ J m$^{-2}$. Photo-oxidation does not occur when the samples are left in darkness, nor is significant oxidation visible when the WS$_2$ is exposed to wavelengths that are not absorbed appreciably – as found in the low-irradiance 760 nm experiment. Taking into consideration that oxidation is only observed upon excitation of an electronic transition, it is likely that no fluence threshold exists, and that oxidation occurs on a probabilistic basis. That is, lower light levels progress the reaction at much slower rates that are difficult to observe using the LSCM or AFM. Still, our study already places severe constraints on the processing and analysis of S-TMD films in ambient conditions, since exposure to typical room light for extended periods (days), or exposure to a single typical scanning PL or Raman spectroscopy measurement (minutes) may cause significant oxidative damage. Furthermore, we expect that these findings will guide the development of new nanofabrication techniques that completely avoid significant oxidation of S-TMDs, as the storage of S-TMDs in darkness or exposure to light that cannot excite electronic transitions should completely prevent progression of the oxidation reaction.

## Methods

 **Chemical Vapor Deposition**. An atmospheric-pressure CVD method was used to grow monolayer WS$_2$ on polished (single side) c-sapphire(0001) substrates (Shinkosha) in a 1-inch quartz tube furnace. The samples were determined to be monolayer from PL and AFM measurements, as presented in the main text. Before beginning the CVD process, the c-sapphire substrates were ultra-sonicated in acetone and ethanol, and then loaded into a 1-inch tube furnace in preparation for oxygen annealing. The samples were oxygen annealed by flowing oxygen at 50 sccm for at least 5 minutes over the sample after which the furnace was then ramped from 30 °C to 1050 °C in 30 minutes, held for 1 hour and then allowed to naturally cool to room temperature. After oxygen annealing, approximately 1.1 ±0.1 g of sulfur powder (≥99.5%, Sigma-Aldrich) and 2 ±0.1 g of WO$_3$ powder (≥99.9%, Sigma-Aldrich) were loaded in separate quartz boats. The sulfur was placed upstream and away from the central part of the furnace heating coils and under external heating coils (controlled by a separate temperature controller), while the WO$_3$ was loaded in the central part of the furnace along with the oxygen annealed substrates. A quartz test tube (approximately 22 mm in diameter) was placed against the substrate quartz boat (with the open-end facing outwards) so as to restrict the flow of reactants (see Supporting Figure S1). This helps increase the partial pressure of reactants over the substrates, thus promoting growth of WS$_2$. The furnace setup is illustrated in Supporting Information Figure S1. Argon was used as the carrier gas and flowed at 200 sccm, and H$_2$ was also added to the gas stream at 8 sccm after purging with argon for 10 minutes. The furnace and external heating coils were then heated to approximately 900 °C and 200 °C from 30 °C in 30 minutes, respectively and held for approximately 5 minutes at these temperatures. After cooling to 800 °C, the furnace lid was opened slightly, so as to rapidly cool the furnace. Between 600 °C and 500 °C, the furnace was fully opened, and the $H_2$ gas turned off. The sulfur was allowed to naturally cool from 200 °C to room temperature. The samples were removed and immediately stored in containers wrapped with aluminum foil and stored in a light-tight box.

**Photoluminescence spectroscopy.** Crystals were analyzed using a WITec 300R spectrometer equipped with a 532 nm laser and power set at approximately 140 µW (as judged by an optical power meter after the 100× objective). 2D confocal PL maps (µ-PL) were taken using the 100×/0.9 (Olympus MPLFLN) objective lens. A Gaussian beam profile was assumed, and the maximum irradiance estimated as $I = 8P/(\pi D^2)$, where $P$ is the laser power and $D$ is the diameter of the beam, estimated as $D = 1.22\lambda/NA \approx 721$ nm, where $\lambda$ is the wavelength in nanometers and $NA$ is the numerical aperture of the lens. The µ-PL map shown in Figs. 2a and 2b took ≈25 minutes to complete (75×75 pixels, ≈ 0.27 seconds integration time per pixel). All PL measurements were under taken at 20 °C ambient temperature and rel. humidity of 40 – 60%.

**Laser scanning confocal microscopy.** Two different LSCM's were used. In Fig 1b, a Keyence VK-X200 system was employed. In all other LSCM figures, an Olympus OLS4100 system was used. These microscopes are also able to take standard optical images, as shown juxtaposed in figures 1b and 1c in the main text, by use of a white LED. The white LED irradiance was calculated for the OLS4100 by first measuring the spectrum of the white LED, whose output was confined to the range 420 – 650 nm. A power meter was then used to measure the optical power exiting the 100×/0.95 objective lens. The responsivity of the power meter's silicon photodetector varies monotonically over this range of wavelengths; hence we were able to put bounds on the power measurement by calibrating the responsivity at two extrema of the spectrum, 420 nm and 650 nm. The power was found to be bounded by ≈0.210±0.070 mW. The field of view of the 100×/0.95 is 128 µm, and the irradiance was then calculated using $I = 4P/(\pi D^2)$. For the blue laser at 405 nm, a power meter measured the maximum output as ≈ 14 µW after the 100×/0.95 objective. The irradiance was then calculated assuming a Gaussian beam, similar to the photoluminescence setup. The time taken per scan varied depending on the magnification used, though the longest time of approximately 50 seconds (corresponding to un-zoomed 100x magnification) was used to calculate the fluence, $H$, in which case the irradiance was multiplied by the total exposure time for each pixel, $t$: $H = 2Pt/(\pi r^2)$. The exposure time was calculated by dividing the total time for the scan (50 seconds) by the total number of pixels (1024×1024). The laser beam diameter is much larger than a pixel, and so would result in the laser 'spilling over' to multiple pixels. To estimate this effect, the area of the laser beam was divided by the area of a pixel to give ≈14. Thus, the calculated time was multiplied by 14 to estimate an upper limit to the true exposure time, yielding ≈ 8.8×10^4 J m^-2. A similar estimate is obtained by dividing the total energy delivered during the complete scan by the total area imaged.

**Atomic force microscopy.** A Bruker Dimension Icon AFM in tapping mode (in ambient and at ≈ 20 °C) was used to obtain the atomic force micrographs. RTESPA-300 tips were used, and analysis was conducted using WSxM v4.0.[67] In Figures 2f, 2g, 2h, the image was flattened, and the contrast enhanced. Bruker's own software was also used to analyze the height of the oxidized triangular islands (using the step height analysis function) and this data can be found in the Supporting Information.

**Controlled low-light irradiance/fluence experiments.** Light sources were a 532 nm laser diode, 650 nm laser diode (measured to be 660 nm, see Supporting Information) and halogen light source (XGY-II) with appropriate bandpass filter at 760 nm ± 10 nm (Thorlabs FB760-10, OD 6+ 200 nm to 719 nm). The optical spectra of the laser sources can be found in the Supporting Information. A custom-built enclosure was used to shroud the experiment in darkness. In the enclosure, the optics used were an iris (Thorlabs ID25), beam steering mirrors, a circular linear variable neutral density filter and a simple convex lens to expand the beam to an appropriate size. Samples were mounted onto a glass slide using double sided sticky tape. Standard samples (usually a sister sample from the same growth run), were kept in darkness for the entirety of the light exposure experiments. The irradiance and fluence for the sources were calculated similarly to the LSCM and PL sections above with the exception that the 760 nm light was assumed to be non-Gaussian (as in the white LED case with the LSCM), i.e $I = 4P/(\pi D^2)$. The diameters were measured as ≈20 mm, ≈12 mm and ≈30 mm  Values for the irradiance were calculated as 3.8 W m$^{-2}$ 14.2 W m$^{-2}$ and 2.2 W m$^{-2}$ for the 532 nm, 660 nm and 760 nm light respectively. Fluence values were then found by multiplying the irradiance by 604800 seconds (seconds in 7 days) to give $2.3\times10^6$, $8.6\times10^6$ and $1.3\times10^6$ J m$^{-2}$ for the 532, 660 and 760 nm light respectively.

**Reflectance / Transmission / Absorptance Measurements and Calculations.** The reflectance and transmission spectra of the WS$_2$ monolayer crystals were measured over 350 – 800 nm using a Perkin Elmer Lambda 1050, equipped with an integrating sphere. Samples grown on c-sapphire (Shinkosha) were mounted on clear glass microscope slides (Sail). Blank sapphire substrates (Shinkosha) mounted on clear glass microscope slides (Sail) from the same batches were also measured. This allowed the absorptance from only the WS$_2$ to be determined in the following way. The absorptance of the substrates, A$_s$ = 1 – T$_s$ – R$_s$, was subtracted from the absorptance of the WS$_2$ and substrate, given as A$_{WS2+S}$ = 1 – T$_{WS2+S}$ – R$_{WS2+S}$, to yield the absorptance from only the WS$_2$ using the formula A$_{WS2}$ = A$_{WS2+S}$ – A$_S$ = (T$_S$ – T$_{WS2+S}$) + (R$_S$ – R$_{WS2+S}$), where T$_i$ and R$_i$ are the measured transmittance and reflectance respectively of the substrate (i = S) and WS$_2$ plus substrate (i = WS2+S). The transmittance, reflectance and absorptance of the WS$_2$ on the substrate and the substrate itself are shown in the Supporting Information.

## Associated Content

Supporting Information is available. Furnace setup is elucidated. The effects of multiple LSCM scans are visualized. Oxidation height analysis with AFM of triangular islands is presented. Details of the edges of WS$_2$ is analyzed further with LSCM and AFM. The spectra of the lasers used in the study are presented, and the transmittance reflectance and absorptance of the WS$_2$ on sapphire on glass and sapphire on glass are presented. Details of FRET and photo-catalysis are given, and possible reaction products relating to the microscopic droplet feature seen in Figure 2g-h are discussed. The effects of heating monolayer WS$_2$ (in the absence of light) is presented. A crystal which underwent minute amounts of oxidation with the 760 nm experiment is presented. Finally, crystals exposed to green light in a nitrogen glove box are compared with ambient air exposed samples.

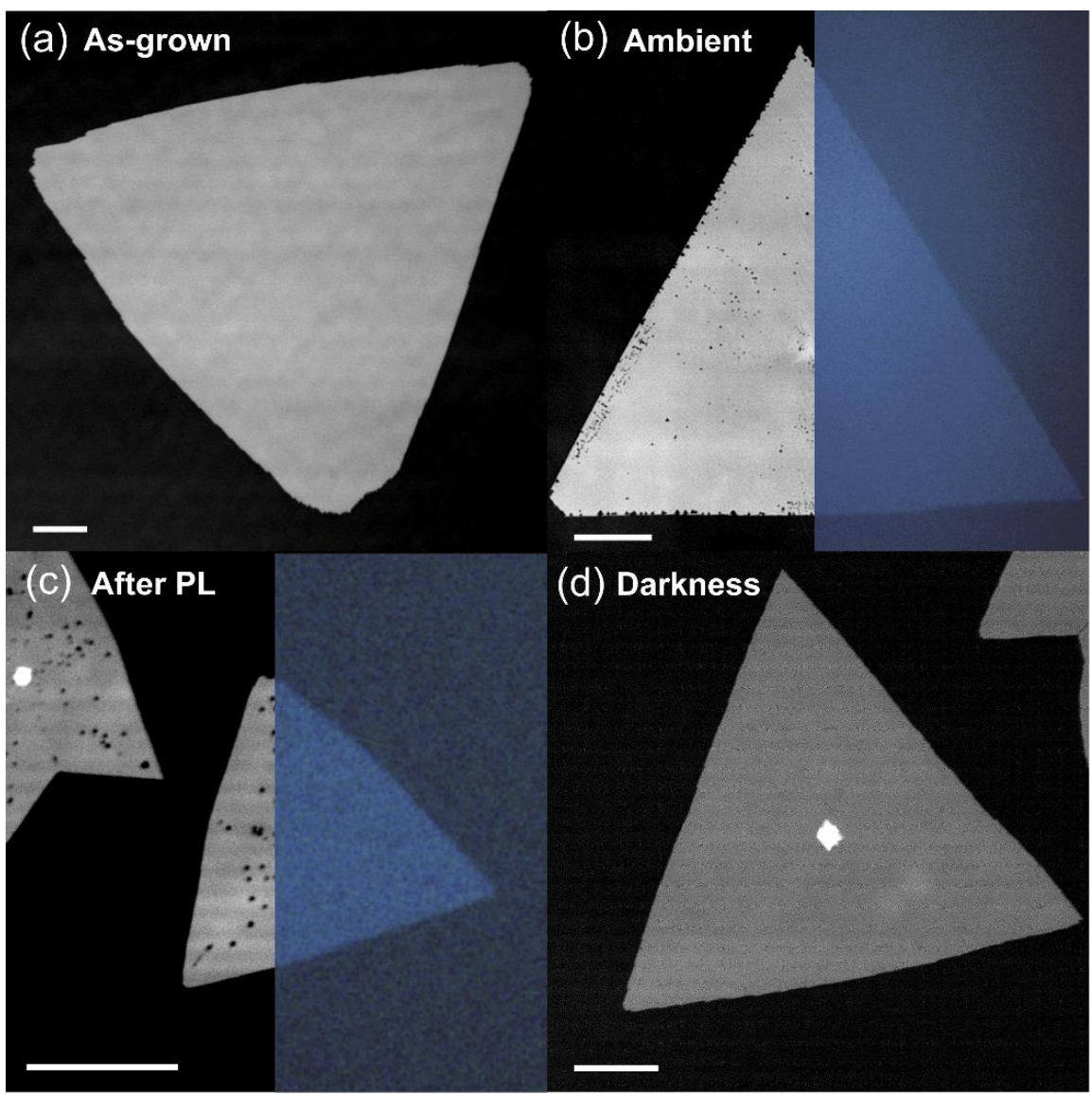

**Figure 1.** Laser scanning confocal micrograph (LSCM) of **(a)** CVD grown monolayer $WS_2$ exposed to minimal amounts of light before imaging with the LSCM approximately 1 month after growth. **(b)** $WS_2$ after approximately 19 days in ambient conditions, juxtaposed with an optical image of the same crystal. **(c)** $WS_2$ after routine photoluminescence spectroscopy, juxtaposed with an optical image of the same crystal and **(d)** WS$_2$ crystals kept in darkness for approximately 10 months, with brief exposure to ambient light. In the middle of this crystal is a crystal seed center, which was common throughout growths. Scale bars in all images are 10 µm.

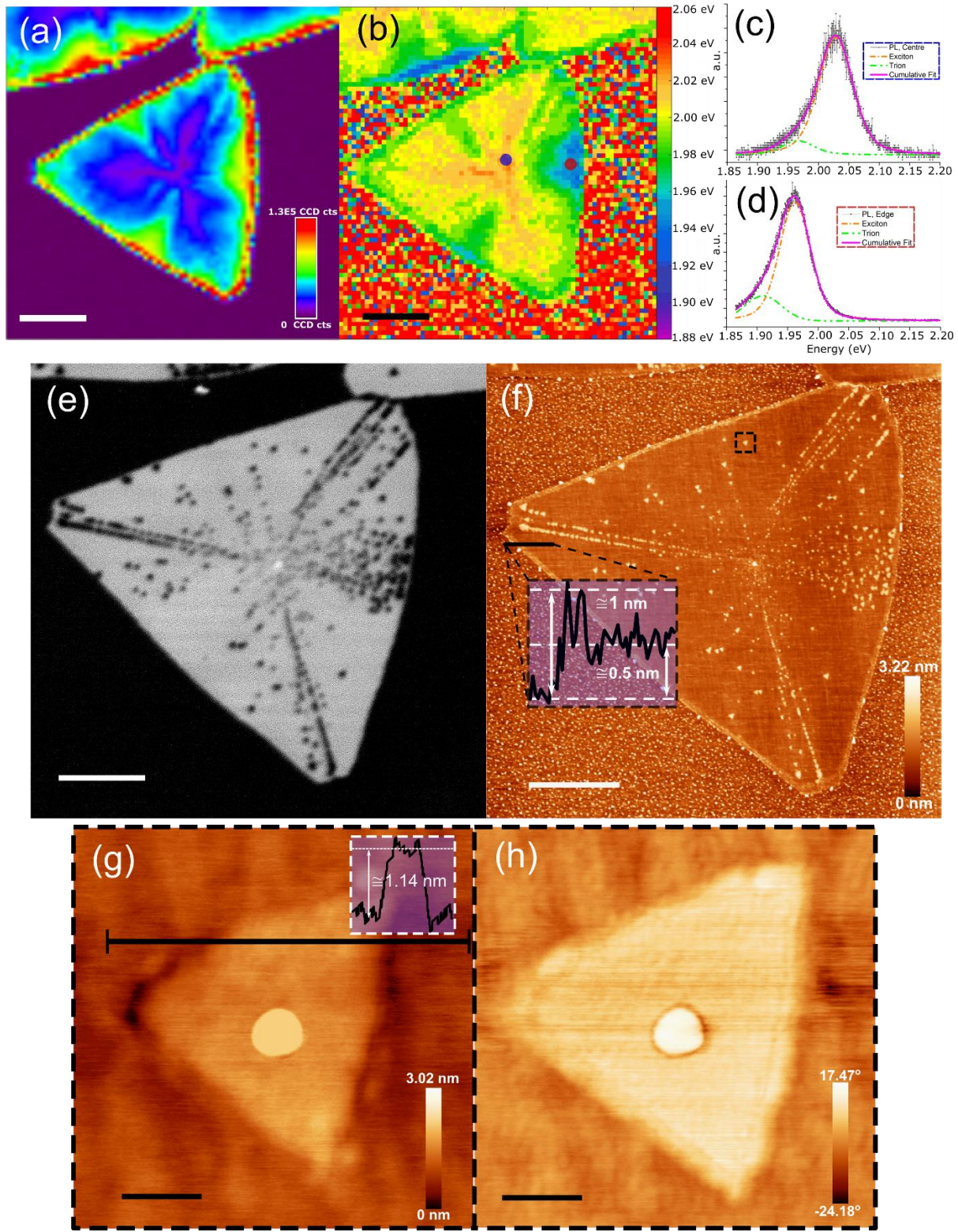

**Figure 2. (a), (b), (c), d)** μ-PL of monolayer WS₂. **(a)** Map showing the intensity in charged

coupled device (CCD) counts (cts) (scale bar = 5 μm) and **(b)** shows the approximate location of the peak in intensity from (a) (scale bar = 5 µm). **(c)** shows the PL spectrum obtained from the central spot (blue) in (b) with exciton and trion peak energies at 2.03±0.01 and 1.97±0.01 eV respectively. **(d)** shows the PL spectrum from the edge region (red spot) with exciton and trion peak energies of 1.96±0.01 and 1.91±0.01 eV respectively. **(e), (f)** LSCM and AFM images showing oxidation of the same monolayer $WS_2$ crystal in Figure 2, after analysis with µ-PL (scale bars = 4 µm). In **(e)**, an LSCM micrograph shows the oxidation as dark spots due to their altered transmittance. The morphology of these oxidation triangles was further investigated using AFM. In **(f)**, the tapping mode AFM height micrograph shows unambiguously that these oxidation zones are raised, triangular and aligned along the stress/strain lines of the crystal. Inset shows the measured height of the crystal. **(g), (h)** Show a zoom in of the boxed area (scale bars = 65 nm) in **(f)** of a triangular oxidation island. The AFM height image in **(g)** shows the height of this island is ≈1.14 nm (inset) and the AFM phase image in **(h)**, shows that the island is different in phase from the surrounding $WS_2$.

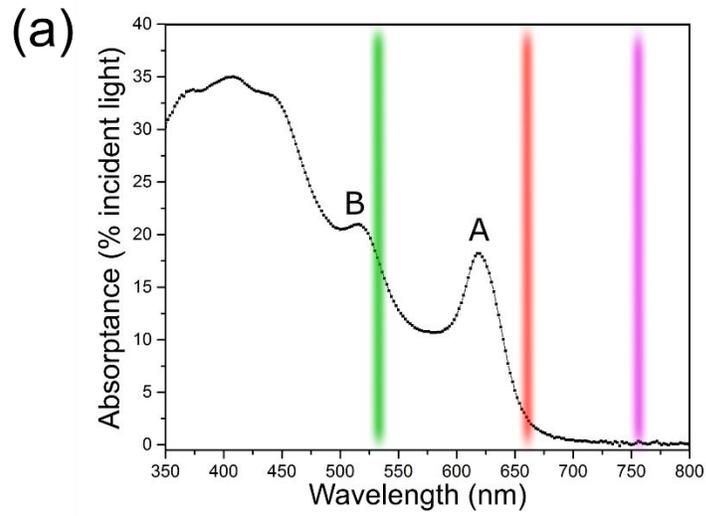

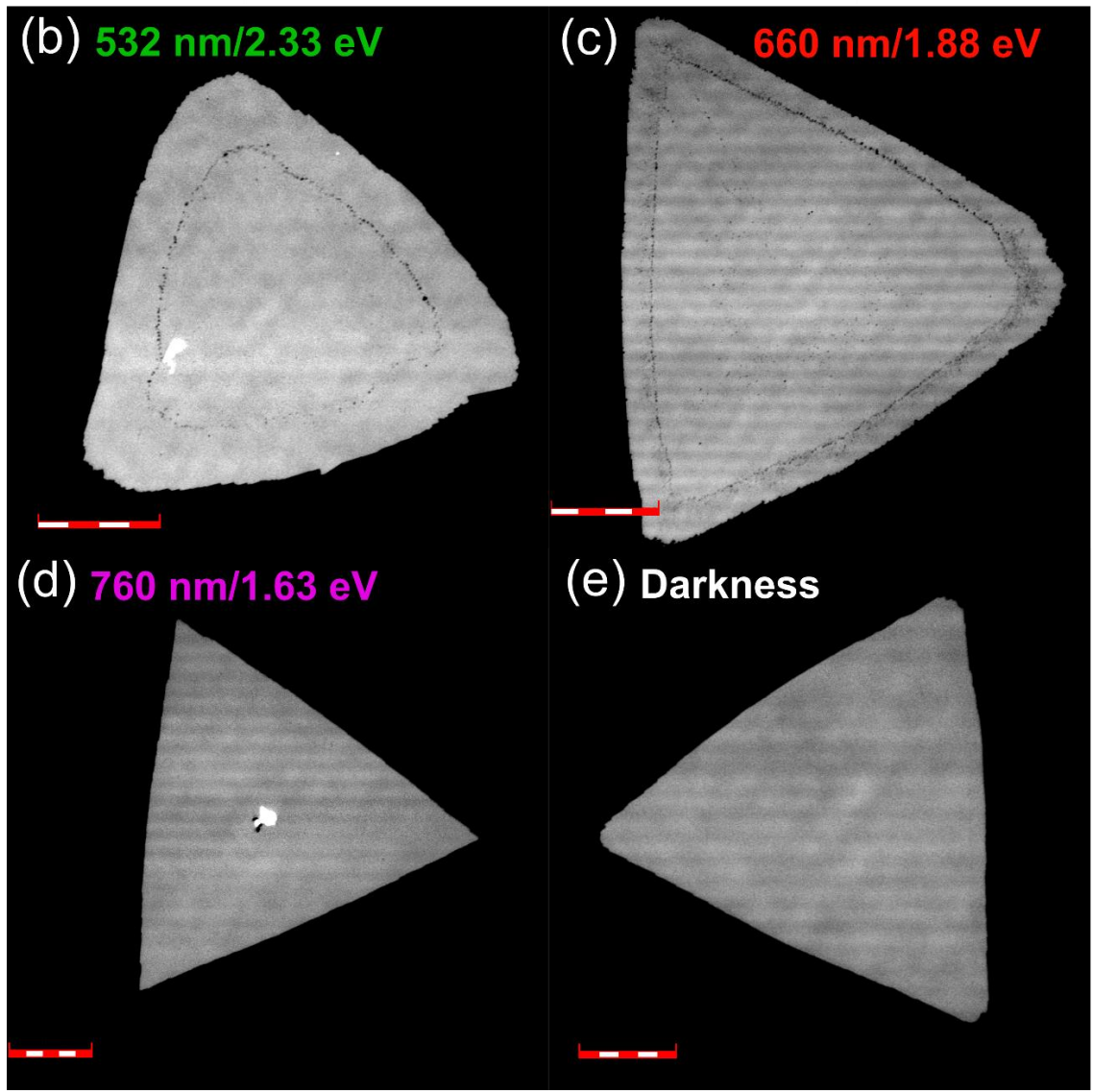

**Figure 3. (a)** Shows the absorbance spectrum of monolayer $WS_2$ taken at $\approx$293 K. Labels denote the 'A' exciton at 619$\pm$2 nm (2.003$\pm$0.006 eV) and 'B' exciton at 516$\pm$2 nm (2.403$\pm$0.009 eV). Color bands correspond to the wavelengths used for the low light exposure experiments, from left to right these are 532 nm (green), 660 nm (red) and 760 nm (purple). **(b), (c), (d)** Low-irradiance light exposure experiments. Laser scanning confocal micrographs of monolayer $WS_2$ after **(b)** 7 days exposure to green laser (532 nm/2.33 eV, 4.2 W m$^{-2}$) scale bar = 20 $\mu$m, **(c)** 7 days of exposure to red light (650 nm/1.91 eV, 14 W m$^{-2}$) scale bar = 20 $\mu$m, **(c)** 7 days of exposure to far-red light (760 nm/1.63 eV, 2.2 W m$^{-2}$) scale bar = 10 $\mu$m. **(e)** A standard sample kept in darkness for 2 weeks for comparison with light exposed samples, scale bar = 10 $\mu$m.

**Table 1.** Summary of wavelengths, irradiances and corresponding fluences, as well as if oxidation was visible with the LSCM for all experiments and analyses. LOW refers to low-irradiance experiments.

| Wavelength (nm) | Experiment /Analysis | Irradiance (W m$^{-2}$) | Fluence (J m$^{-2}$) | Oxidation Visible? |
|---|---|---|---|---|
| 415– 700 (white light from LSCM) | LSCM | $1.6\times10^4$ | $1.6\times10^5$ | No |
| 405 (blue) | LSCM | $1.3\times10^8$ | $8.8\times10^4$ | No |
| 515 (green) | LED* | 3 | $1.8\times10^6$ | Yes |
| 532 (green) | LOW | 3.8 | $2.3\times10^6$ | Yes |
| 532 (green) | $\mu$-PL | $6.9\times10^8$ | $1.8\times10^8$ | Yes |
| 660 (red) | LOW | 14.2 | $8.6\times10^6$ | Yes |
| 760 (far red) | LOW | 2.2 | $1.3\times10^6$ | No** |

*See Supporting Information Section 10 for details.

**Some crystals had signs of small amounts of oxidation. See Results and Supporting Information.

TOC graphic

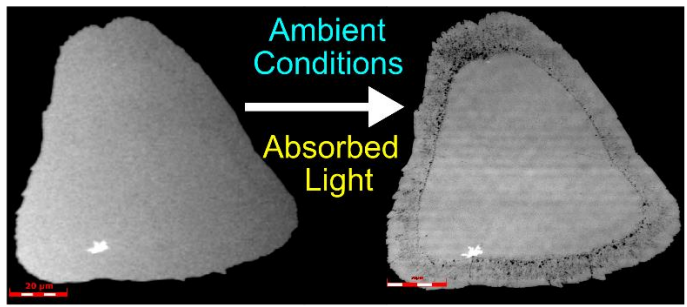

AUTHOR INFORMATION

**Corresponding Authors**

**\*Email (J. C. Kotsakidis): Jimmy.Kotsakidis@monash.edu**

**\*Email (M. S. Fuhrer): Michael.Fuhrer@monash.edu**

**Author Contributions**

J.C.K and M.C made the first observation with the LSCM, which was later developed further by J.C.K with the help of D.K.G, M.S.F, A.L.V.P and Q.Z. J.C.K grew most of the samples, with some samples grown by Q.Z. AFM was conducted by J.C.K and Q.Z. PL was conducted by J.C.K. Low light irradiance/fluence laser experiments were conceived by J.C.K. and A.L.V. P and carried out by J.C.K. Optical absorptance/transmittance/reflectance was conducted by J.C.K and A.C. J.C.K wrote the manuscript with significant editorial contributions from M.S.F, D.K.G and A.L.V.P. The manuscript was written through contributions of all authors. All authors have given approval to the final version of the manuscript.

**Funding Sources**

J.C.K and M.S.F acknowledge funding support from the Australian Research Council (DP150103837). D.K.G and M.C acknowledge funding from the Office of Naval Research (ONR). A.L.V.P acknowledges the A.L.V.P. acknowledges the Ministerio de Economia y Competitividad (MINECO) projects FOS2015-67367-C2-1-P, PGC2018-093291-B-I00 and Comunidad de Madrid project NMAT2d P2018/NMT-4511 for financial support. Q.Z acknowledges funding support from the Australian Research Council (DE190101249).

**Notes**

The authors declare no competing financial interests.

## ACKNOWLEDGMENT

J.C.K would like to thank A. Chesman for transmittance/reflectance/absorptance measurements and useful discussions, F. Shanks for use of the WITec PL/Raman system, as well as useful discussions, M. Greaves for time on the LSCM, S. Bhattacharyya for assistance with the glovebox and S. Johnstone for assistance with equipment in the optical laboratory at Monash University. J.C.K also acknowledges the hospitality of the U.S. Naval Research Laboratory (NRL) in which the initial investigations were performed. This work was performed in part at the Melbourne Centre for Nanofabrication (MCN) in the Victorian Node of the Australian National Fabrication Facility (ANFF).

## ABBREVIATIONS

AFM, Atomic Force Microscopy ; CVD, Chemical Vapor Deposition; PL, Photoluminescence ; μ-PL, confocal micro-photoluminescence (mapping) spectroscopy; LSCM, Laser scanning confocal microscope ; FRET, Förster Resonance Energy Transfer.